# Synthesis of Fault Tolerant Reversible Logic Circuits

Md. Saiful Islam[1], M. M. Rahman[1], Zerina Begum[1]
[1]Institute of Information Technology
University of Dhaka, Dhaka-1000, Bangladesh
e-mail: saifulit@univdhaka.edu,
mus_mahbub@hotmail.com, zerin@univdhaka.edu

Mohd. Zulfiquar Hafiz[1], Abdullah Al Mahmud[2]
[2]Dept. of Computer Science and Engineering
Ahsanullah University of Science and Technology
Dhaka-1208, Bangladesh
e-mail: jewel@univdhaka.edu, aamrubel@yahoo.com

*Abstract*—Reversible logic is emerging as an important research area having its application in diverse fields such as low power CMOS design, digital signal processing, cryptography, quantum computing and optical information processing. This paper presents a new 4*4 universal reversible logic gate, IG. It is a parity preserving reversible logic gate, that is, the parity of the inputs matches the parity of the outputs. The proposed parity preserving reversible gate can be used to synthesize any arbitrary Boolean function. It allows any fault that affects no more than a single signal readily detectable at the circuit's primary outputs. Finally, it is shown how a fault tolerant reversible full adder circuit can be realized using only two IGs. It has also been demonstrated that the proposed design offers less hardware complexity and is efficient in terms of gate count, garbage outputs and constant inputs than the existing counterparts.

## I. Introduction

Power dissipation is an important factor in VLSI design. Combinational logic circuits dissipate heat in an order of $kT \ln 2$ joules for every bit of information that is lost, where $k$ is the Boltzman constant and $T$ is the operating temperature [1]. Information is lost when the input vector can not be uniquely recovered from its output vectors. Reversible logic circuits naturally take care of heating since in a reversible logic every input vector can be uniquely recovered from its output vectors and therefore no information is lost. According to [2] zero energy dissipation would be possible only if the network consists of reversible gates. Thus reversibility will become an essential property in future circuit design.

Synthesis of reversible logic circuits differs from the combinational one in many ways [3]. Firstly, in reversible circuit there should be no fan-out, that is, each output will be used only once. Secondly, for each input pattern there should be a unique output pattern. Finally, the resulting circuit must be acyclic. Any reversible gate performs the permutation of its input patterns only and realizes the functions that are reversible. If a reversible gate has $k$ inputs, and therefore $k$ outputs, then we call it a $k*k$ reversible gate. Any reversible circuit design includes only the gates that are reversible. In a reversible circuit, the outputs that are not used as primary outputs are called garbages and the input lines that are set to constants are termed as constant inputs. An efficient design should keep the number of garbage outputs to minimum.

Parity checking is one of the widely used error detection mechanisms in digital logic and data communication systems. This is because most of the arithmetic functions is not parity preserving. If the parity of the input data is maintained throughout the computation, no intermediate checking would be required [4]. A sufficient requirement for parity preservation of a reversible circuit is that each gate be parity preserving [4]. Thus, we need parity preserving reversible logic gates to construct parity preserving reversible circuits. This paper presents a new 4*4 parity preserving logic gate, IG. It is parity preserving, that is, the parity of the inputs matches the parity of the outputs. IG is universal in the sense that it can be used to synthesize any arbitrary Boolean function. It is shown that a fault tolerant reversible full adder circuit can be realized using only two IGs. The presented design does not produce any unnecessary garbage outputs. Minimizing number garbage outputs are the major concern in reversible logic design [3]. The presented fault tolerant full adder block can be used to realize other arithmetic circuit in nanotechnology such as ripple carry adder, carry look-ahead adder, carry-skip logic, and multiplier/divisors.

## II. Reversible Logic Gates

### A. Basic Reversible Gates

There exist many reversible gates in the literature. Among them 2*2 Feynman gate (FG) [6], depicted in Fig. 1a, 3*3 Peres gate (PG) [7], depicted in Fig. 1b, 3*3 Toffoli gate (TG) [8], depicted in Fig. 1c and 3*3 Fredkin gate (FRG) [9], depicted in Fig. 1d have been studied extensively. Because of their simplicity and quantum realization cost there are design approaches and tools that incorporate them separately or in combination with each other [3] [5].

### B. Parity Preserving Reversible Gates

Fault tolerance is the property that enables a system to continue operating properly in the event of the failure of some its components. If the system itself made of fault tolerant components, then the detection and correction of faults become easier and simple. In communication and many other systems, fault tolerance is achieved by parity. Therefore, parity preserving reversible circuits will be the future design trends to the development of fault tolerant reversible systems in nanotechnology. And a gating network will be parity preserving if its individual gate is parity preserving [4].

A few parity preserving logic gates have been proposed in the literature. Among them 3*3 Feynman Double gate (F2G) [4] depicted in Fig. 2a and 3*3 Fredkin gate (FRG) [9] depicted in Figure 2b are one-through gates, which means one of the inputs is also output. Recently Haghparast and Navi [10] have proposed a new 3*3 parity preserving reversible gate, namely New Fault Tolerant gate (NFT) depicted in Figure 2c.

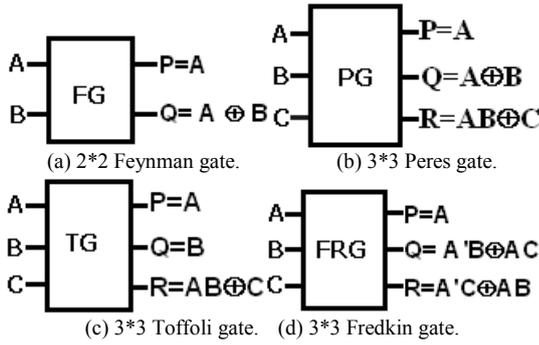

(a) 2*2 Feynman gate.  (b) 3*3 Peres gate.

(c) 3*3 Toffoli gate.  (d) 3*3 Fredkin gate.

Figure 1. Some well-known reversible logic gates.

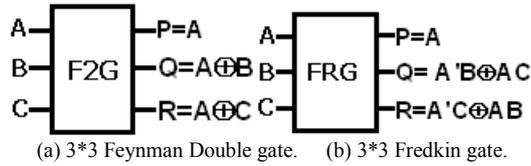

(a) 3*3 Feynman Double gate.  (b) 3*3 Fredkin gate.

(c) 3*3 NFT gate

Figure 2. Some parity preserving reversible gates.

From Table 1, 2, and 3 we can see that the gates FRG, F2G and NFT are parity preserving since they satisfy A⊕B⊕C=P⊕Q⊕R. And any k*k reversible logic gate where the EX-OR of the inputs matches the EX-OR of the outputs will be parity preserving.

*C. A New 4*4 Parity Preserving Reversible Gate*

This paper presents a new 4*4 parity preserving reversible gate, IG, depicted in Fig. 3. The gate is one-through, which means one of the input variables is also output. The corresponding truth table of the gate is shown in Table 4. It can be verified from the truth table that the input pattern corresponding to particular output pattern can be uniquely determined. The proposed reversible IG is parity preserving. This is readily verified by comparing the input parity A⊕B⊕C⊕D to the output parity P⊕Q⊕R⊕S.

The newly proposed IG gate is universal in the sense that it can be used for implementing arbitrary Boolean functions as shown in Fig. 4.

*D. Design of Parity Preserving TG Circuit*

Toffoli gate [8], depicted in Fig.1c, is one of the most important and useful gate in reversible logic circuit synthesis. Therefore implementation of parity preserving Toffoli gate is essential. A parity preserving reversible TG circuit is presented in [4], which is shown in Fig. 5. The circuit requires three reversible gates (one is FRG and two are F2G gates) and produces two garbage outputs. Another parity preserving reversible TG circuit is presented in [10], which is shown in Fig. 6. The circuit requires two reversible gates (one is NFT and one is F2G gate) and produces one garbage output.

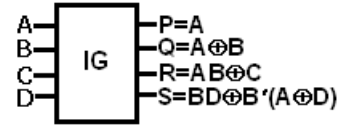

Figure 3. A new 4*4 parity preserving reversible gate IG.

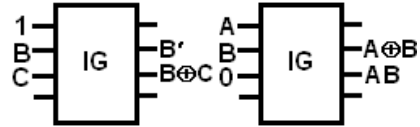

(a) IG as Inverter and EX-OR. (b) IG as AND and EX-OR.

(c) IG as EXOR, EX-NOR and OR

Figure 4. Proposed IG gate can implement all Boolean functions.

TABLE I. TRUTH TABLE OF PARITY PRESERVING FREDKIN GATE

| A | B | C | P | Q | R |
|---|---|---|---|---|---|
| 0 | 0 | 0 | 0 | 0 | 0 |
| 0 | 0 | 1 | 0 | 0 | 1 |
| 0 | 1 | 0 | 0 | 1 | 0 |
| 0 | 1 | 1 | 0 | 1 | 1 |
| 1 | 0 | 0 | 1 | 0 | 0 |
| 1 | 0 | 1 | 1 | 1 | 0 |
| 1 | 1 | 0 | 1 | 0 | 1 |
| 1 | 1 | 1 | 1 | 1 | 1 |

TABLE II. TRUTH TABLE OF PARITY PRESERVING F2G GATE

| A | B | C | P | Q | R |
|---|---|---|---|---|---|
| 0 | 0 | 0 | 0 | 0 | 0 |
| 0 | 0 | 1 | 0 | 0 | 1 |
| 0 | 1 | 0 | 0 | 1 | 0 |
| 0 | 1 | 1 | 0 | 1 | 1 |
| 1 | 0 | 0 | 1 | 1 | 1 |
| 1 | 0 | 1 | 1 | 1 | 0 |
| 1 | 1 | 0 | 1 | 0 | 1 |
| 1 | 1 | 1 | 1 | 0 | 0 |

TABLE III. TRUTH TABLE OF PARITY PRESERVING NFT GATE

| A | B | C | P | Q | R |
|---|---|---|---|---|---|
| 0 | 0 | 0 | 0 | 0 | 0 |
| 0 | 0 | 1 | 0 | 1 | 0 |
| 0 | 1 | 0 | 1 | 0 | 0 |
| 0 | 1 | 1 | 1 | 0 | 1 |
| 1 | 0 | 0 | 1 | 1 | 1 |
| 1 | 0 | 1 | 1 | 1 | 0 |
| 1 | 1 | 0 | 0 | 0 | 1 |
| 1 | 1 | 1 | 0 | 0 | 1 |

This paper presents a new parity preserving reversible TG circuit, shown in Fig.7. The circuit requires only two reversible gates (one is FRG and one is F2G) and produces only one garbage output.

*E. Evaluation of the Proposed Parity Preserving TG Circuit*

The proposed parity preserving reversible TG circuit is more efficient than the existing circuits presented in [4], [10]. Let

α = A two input EXOR gate calculation
β = A two input AND gate calculation
δ = A NOT gate calculation
T = Total logical calculation

TABLE IV. TRUTH TABLE OF PARITY PRESERVING IG GATE

| A | B | C | D | P | Q | R | S |
|---|---|---|---|---|---|---|---|
| 0 | 0 | 0 | 0 | 0 | 0 | 0 | 0 |
| 0 | 0 | 0 | 1 | 0 | 0 | 0 | 1 |
| 0 | 0 | 1 | 0 | 0 | 0 | 1 | 0 |
| 0 | 0 | 1 | 1 | 0 | 0 | 1 | 1 |
| 0 | 1 | 0 | 0 | 0 | 1 | 0 | 0 |
| 0 | 1 | 0 | 1 | 0 | 1 | 0 | 1 |
| 0 | 1 | 1 | 0 | 0 | 1 | 1 | 0 |
| 0 | 1 | 1 | 1 | 0 | 1 | 1 | 1 |
| 1 | 0 | 0 | 0 | 1 | 0 | 0 | 1 |
| 1 | 0 | 0 | 1 | 1 | 0 | 0 | 1 |
| 1 | 0 | 1 | 0 | 1 | 1 | 1 | 1 |
| 1 | 0 | 1 | 1 | 1 | 1 | 1 | 0 |
| 1 | 1 | 0 | 0 | 1 | 0 | 1 | 0 |
| 1 | 1 | 0 | 1 | 1 | 0 | 1 | 1 |
| 1 | 1 | 1 | 0 | 1 | 0 | 0 | 0 |
| 1 | 1 | 1 | 1 | 1 | 0 | 0 | 1 |

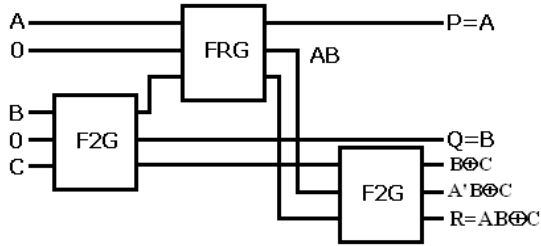

Figure 5. Existing TG with parity preservation, presented in [4].

For [4] the Total logical calculation is: $T=6\alpha+4\beta+2\delta$, for [10] the Total logical calculation is: $5\alpha+3\beta+2\delta$, and for our proposed parity preserving reversible TG circuit, the Total logical calculation is: $4\alpha+4\beta+\delta$. Therefore, the presented design offers less hardware complexity than the existing counterparts.

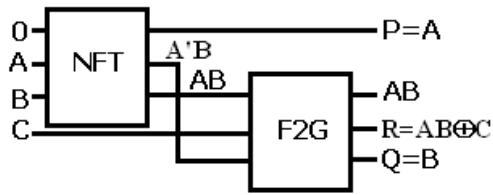

Figure 6. Existing TG with parity preservation, presented in [10].

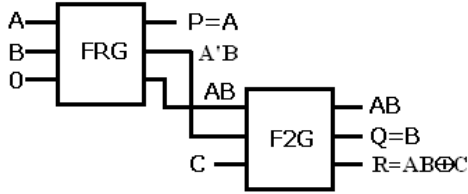

Figure 7. Proposed parity preserving TG implementation.

### III. FAULT TOLERANT FULL ADDER CIRCUIT

Reversible logic implementation of full adder circuit has been studied by several authors in the literature [3-4][12]. It has been demonstrated in [3] that a reversible full adder circuit can be realized with at least two garbage outputs and one constant input. This requirement is not the same for fault tolerant reversible full adder circuit. Because in a fault tolerant full adder circuit the input parity must matches the parity of the outputs. This section first establishes the minimum number of garbage outputs and constant inputs required to design a fault tolerant reversible full adder circuit. Then proposes a new realization of fault tolerant reversible full adder circuit using the newly proposed IG gates.

**Theorem 2:** A fault tolerant reversible full adder circuit can be realized with at least three garbage outputs and two constant inputs.

*Proof*: The full adder circuit output equations $S=A\oplus B\oplus C_{in}$ and $C_{out}=(A\oplus B)C_{in}\oplus AB$ produce the same output S=1 and $C_{out}$=0, for the three distinct input combinations A=0, B=0, $C_{in}$=1; A=0, B=1, $C_{in}$=0; and A=1, B=0, $C_{in}$=0. The parity of the input vector matches the parity of the corresponding output vector. To separate all repeated values of outputs S and $C_{out}$ as well as keeping their parity unchanged we need at least three garbage outputs. Thus the total number of outputs is 2+3=5. Now since in a reversible circuit the number of inputs must be equal to the number of outputs and there are three inputs in a full adder circuit A, B and $C_{in}$, the other two inputs need to be constant inputs. ∎

TABLE V. INPUT COMBINATIONS THAT PRODUCE THE SAME OUTPUT COMBINATIONS IN FULL ADDER CIRCUIT (SHOWN SHADED)

| Input | | | | | Output | | | | |
|---|---|---|---|---|---|---|---|---|---|
| A | B | Cin | C1 | C2 | S | Cout | G1 | G2 | G3 |
| 0 | 0 | 1 | 0 | 0 | 1 | 0 | 0 | 0 | 0 |
| 0 | 1 | 0 | 0 | 0 | 1 | 0 | 0 | 1 | 1 |
| 1 | 0 | 0 | 0 | 0 | 1 | 0 | 1 | 0 | 1 |

There are two fault tolerant reversible full adder circuits in the literature [11], [12]. The fault tolerant full adder circuit in [11] requires six parity preserving reversible gates (two FRGs and four F2Gs) and the fault tolerant full adder circuit in [12] uses four FRGs. This paper presents a new design of fault tolerant reversible full adder circuit that uses only two IGs, depicted in Fig. 8. It requires only two clock cycles.

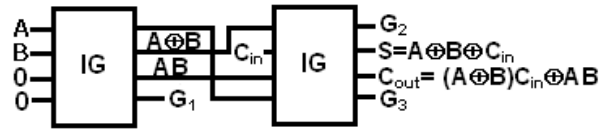

Figure 8. Proposed fault tolerant reversible full adder circuit.

#### A. Evaluation of the Proposed Full Adder Circuit

Evaluation of the proposed circuit can be comprehended easily with the help of the comparative results given in Table 6.

*1) Hardware Complexity*: One of the main factors of a circuit is its hardware complexity. It can be proved that the proposed circuit is better than the existing approaches in terms of hardware complexity. For [12] the Total logical calculation is: $T=8\alpha+16\beta+4\delta$, for [11] the Total logical calculation is: $12\alpha+8\beta+2\delta$, and for our proposed parity preserving reversible full adder circuit, the Total logical calculation is: $8\alpha+6\beta+2\delta$. Therefore, the hardware complexity of the proposed parity preserving reversible full adder circuit is less than the existing counterparts.

TABLE VI. COMPARATIVE RESULTS OF DIFFERENT FAULT TOLERANT FULL ADDER CIRCUITS

|  | No. of Gates Used | No. of Clock Cycles | No. of Garbage Outputs | No. of Constant Inputs | Total Logical Calculation |
|---|---|---|---|---|---|
| Proposed Circuit | 2 IGs=2 | 2 | 3 | 2 | $8\alpha+6\beta+2\delta$ |
| Existing Circuit[12] | 4 FRGs =4 | 4 | 3 | 2 | $8\alpha+16\beta+4\delta$ |
| Existing Circuit[11] | 2FRGs+4 F2Gs=6 | 6 | 6 | 5 | $12\alpha+8\beta+2\delta$ |

*2) Garbage Outputs*: Garbage output refers to the output of the reversible gate that is not used as a primary output or as input to other gates [3]. One of the other major constraints in designing a reversible logic circuit is to lessen number of garbage outputs. Our proposed parity preserving reversible full adder circuit produces only three garbage outputs which are equal to the design in [12] and this is the minimum as proved earlier in this paper, but the design in [11] produces six garbage outputs. So, it can be stated that the proposed design approach is better than all the existing counterparts in terms of number of garbage outputs.

*3) Constant Inputs*: Number of constant inputs is one of the other main factors in designing a reversible logic circuit. The input that is added to an n*k function to make it reversible is called constant input [3]. The proposed parity preserving reversible full adder circuit requires only two constant inputs that are equal to the design in [12] and this is the minimum theoretically, but the design in [11] requires 5 constant inputs. So, it can be stated that the proposed design approach is better than all the existing designs in terms of number of constant inputs.

From the above discussion we can conclude that the proposed fault tolerant reversible full adder circuit is better than all the existing counterparts.

### B. Ripple Carry Adder

The full adder is the basic building block in a ripple carry adder. The reversible ripple carry adder using the fault tolerant full adder (FTFA) is shown in Fig. 9 which is obtained by cascading the full adders in series. The output expressions for a ripple carry adder are:

$$S_i = A \wedge B \wedge C_i \quad (1)$$
$$C_{i+1} = (A \wedge B) \cdot C_i \wedge AB \quad (i=0, 1, 2\ldots) \quad (2)$$

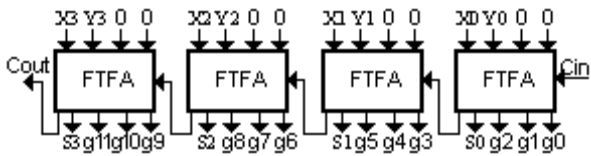

Figure 9. Ripple carry adder using the proposed FTFA.

*Evaluation of the Proposed Ripple Carry Adder:* It can be inferred from Fig. 8 and Fig. 9 that for N bit addition; the proposed ripple carry adder architecture uses only 2N reversible IG gates and produces only 3N garbage outputs. Table 7 shows the results that compare the proposed ripple carry adder with those designed using full adders of [11-12]. It is observed that the proposed circuit is better than existing ones both in terms of gate count and garbage outputs.

TABLE VII. COMPARATIVE RESULTS OF THE REVERSIBLE RIPPLE CARRY ADDER

|  | No. of Reversible Gates Used | No. of Garbage Outputs Produced | No. of Constant Inputs Required | Unit Delay |
|---|---|---|---|---|
| Proposed Circuit | 2N | 3N | 2N | 2N |
| Existing Circuit [12] | 4N | 3N | 2N | 4N |
| Existing Circuit [11] | 6N | 6N | 5N | 6N |

## IV. CONCLUSION

This paper presents a new 4*4 parity preserving reversible gate called IG gate and demonstrates its universality by realizing all possible Boolean functions. A novel fault tolerant reversible full adder circuit using the proposed IG gates has also been presented and optimized in terms of gate count, number of garbage outputs and constant inputs. Reversible logic implementation of optimized fault tolerant N-bit ripple carry adder has also been presented. The presented adder architectures using the proposed reversible gate offer less hardware complexity and optimized in terms of area and power consumption.

REFERENCES

[1] R. Landauer, "Irreversibility and heat generation in the computational process", IBM J. Res. Develop., vol. 5, pp. 183-191, 1961.
[2] C. H. Bennet, "Logical reversibility of computation", IBM J. Res. Develop., vol. 17, no. 6, pp. 525-532, 1973.
[3] M. S. Islam, and M. Rafiqul Islam, "Minimization of reversible adder circuits", Asian Journal of Information Technology, vol. 4, no. 12, pp. 1146-1151, 2005.
[4] B. Parhami, "Fault tolerant reversible circuits", in Proceedings of 40th Asimolar Conf. Signals, Systems, and Computers, Pacific Grove, CA, pp. 1726-1729, October 2006.
[5] D. Maslov, G. W. Dueck, and D. M. Miller, "Synthesis of Fredkin-Toffoli reversible networks," IEEE Trans. VLSI Systems, vol. 13, no. 6, pp. 765-769, 2005.
[6] R. Feynman, "Quantum mechanical computers", Optical News, vol. 11, 1985, pp. 11-20.
[7] A. Peres, "Reversible logic and quantum computers", Physical Review: A, vol. 32, no. 6, pp. 3266-3276, 1985.
[8] T. Toffoli, "Reversible computing", In Automata, Languages and Programming, Springer-Verlag, pp. 632-644, 1980.
[9] E. Fredkin and T. Toffoli, "Conservative logic", Intl. Journal of Theoretical Physics, pp. 219-253, 1982.
[10] M. Haghparast and K. Navi, "A novel fault tolerant reversible gate for nanotechnology based systems", Am. J. of App. Sci., vol. 5, no.5, pp. 519-523, 2008.
[11] M. Haghparast and K. Navi, "Design of a novel fault tolerant reversible full adder for nanotechnology based systems", World App. Sci. J., vol. 3, no. 1, pp. 114-118, 2008.
[12] J. W. Bruce, M. A. Thornton, L. Shivakumaraiah, P.S. Kokate, X. Li, "Efficient adder circuits based on a conservative reversible logic gates", In Proceedings of IEEE Computer Society Annual Symposium on VLSI, Pittsburg, PA, pp. 83-88, 2002.